\documentclass[aps,prb,reprint,superscriptaddress]{revtex4-1}

\usepackage{graphicx}
\usepackage{bm}
\usepackage{amsmath}
\usepackage{epstopdf}
\usepackage{multirow}

\begin{document}

\preprint{APS/PRB}

\title{Magnetic structures in the rich magnetic phase diagram of Ho$_2$RhIn$_8$}

\author{Petr \v{C}erm\'{a}k}
\affiliation{%
J{\"u}lich Centre for Neutron Science JCNS, Forschungszentrum J{\"u}lich GmbH, Outstation at MLZ, Lichtenbergstra{\ss}e 1, 85747 Garching, Germany
}%
\affiliation{%
Charles University, Faculty of Mathematics and Physics, Department
of Condensed Matter Physics, Ke~Karlovu~5, 121 16 Prague 2, The Czech Republic
}%
\email{p.cermak@fz-juelich.de}

\author{Karel Proke\v{s}}
\affiliation{%
Helmholtz-Zentrum Berlin f\"ar Materialien und Energie, Hahn-Meitner-Platz 1, Berlin 14109, Germany
}%

 \author{Bachir Ouladdiaf}
\affiliation{%
Institut Laue Langevin, 6 rue Jules Horowitz, BP156, 38042 Grenoble Cedex 9, France
}%

\author{Martin Boehm}
\affiliation{%
 Institut Laue Langevin, 6 rue Jules Horowitz, BP156, 38042 Grenoble Cedex 9, France
}%

\author{Marie Kratochv\'{\i}lov\'a}
\affiliation{%
Charles University, Faculty of Mathematics and Physics, Department
of Condensed Matter Physics, Ke~Karlovu~5, 121 16 Prague 2, The Czech Republic
}%

\author{Pavel Javorsk\'y}
\affiliation{%
Charles University, Faculty of Mathematics and Physics, Department
of Condensed Matter Physics, Ke~Karlovu~5, 121 16 Prague 2, The Czech Republic
}%

\date{\today}

\begin{abstract}
The magnetic phase diagram of the tetragonal Ho$_2$RhIn$_8$ compound has similar features to many related systems, 
revealing a zero magnetic field AF1 and a field-induced AF2 phases.
Details of the magnetic order in the AF2 phase were not reported yet for any of the related compounds.
In addition, only the Ho$_2$RhIn$_8$ phase diagram contains a small region of the incommensurate zero-field AF3 phase.
We have performed a number of neutron diffraction experiments on single crystals of Ho$_2$RhIn$_8$ using several diffractometers
including experiments in both horizontal and vertical magnetic fields up to 4 T.
We present details of the magnetic structures in all magnetic phases of the rich phase diagram of Ho$_2$RhIn$_8$. 
The Ho magnetic moments point along the tetragonal $c$ axis in every phase.
The ground-state AF1 phase is characterized by propagation vector $\textbf{k}$ = (1/2, 0, 0).
The more complex ferrimagnetic AF2 phase is described by four propagation vectors
$\textbf{k}_{0}$ = (0, 0, 0), $\textbf{k}_{1}$ = (1/2, 0, 0), $\textbf{k}_{2}$ = (0, 1/2, 1/2), $\textbf{k}_{3}$ = (1/2, 1/2, 1/2).
The magnetic structure in the AF3 phase is incommensurate with $\textbf{k}_{AF3}$ = (0.5, $\delta$, 0).
Our results are consistent with theoretical calculations based on crystal field theory.
\end{abstract}


\maketitle

\section{Introduction}
The heavy-fermion compounds have been attracting the scientific interest for almost two decades
due to presence of magnetic order and unconventional superconductivity induced in the vicinity
of the magnetic quantum critical point \cite{QCP-Thompson}.
Both phenomena can coexist in broad ranges of the temperature-pressure-substitution phase diagrams \cite{Review-Pfleiderer}. 
Compounds with general composition $R_{n}T_{m}X_{3n+2m}$ (where $R$ = rare earth, $T$ = transition metal, $X$ = In or Ga, $n$ and $m$ are integers)
provide, however, another possibility of tuning the system towards quantum criticality due to their layered structure. 
The structural dimensionality plays a crucial role in the character of a magnetic quantum critical point \cite{QCP-Si,QCP-Custers}.
In general, the crystal structure consists of rare-earth ($n$) and transition metal ($m$) layers surrounded by cages of indium atoms.
Different sequences of these layers along tetragonal $c$ axis is reflected in various tetragonal structures with rich ground state properties.
In terms of anisotropy, the "218" structure (e.g. Ce$_2$RhIn$_8$ \cite{CeRh218-SC} or Ce$_2$PdIn$_8$ \cite{Ce2PdIn8,Ce2PdIn8-comment}) 
lies in between the well known quasi-2D "115" structure 
(e.g. CeCoIn$_5$ \cite{CeCo115-SC}, CeRhIn$_5$ \cite{CeRhIn5}) and the cubic CeIn$_3$ structure.
Recently, this "dimensionality row" was enriched by discovering "127" compounds (CePt$_2$In$_7$ \cite{CePt2In7})
and new structure types "3-1-11" and "5-2-19" \cite{Ce3111-Kacz,Ce3111-Krat}
which represent an intermediate step between CeIn$_3$ and the "218" compounds.
As the superconductivity is observed only in cerium compounds, 
studies of non-Kondo materials have a crucial importance for understanding the magnetism in these systems.

\begin{figure}
\resizebox{0.5\textwidth}{!}{%
 \includegraphics{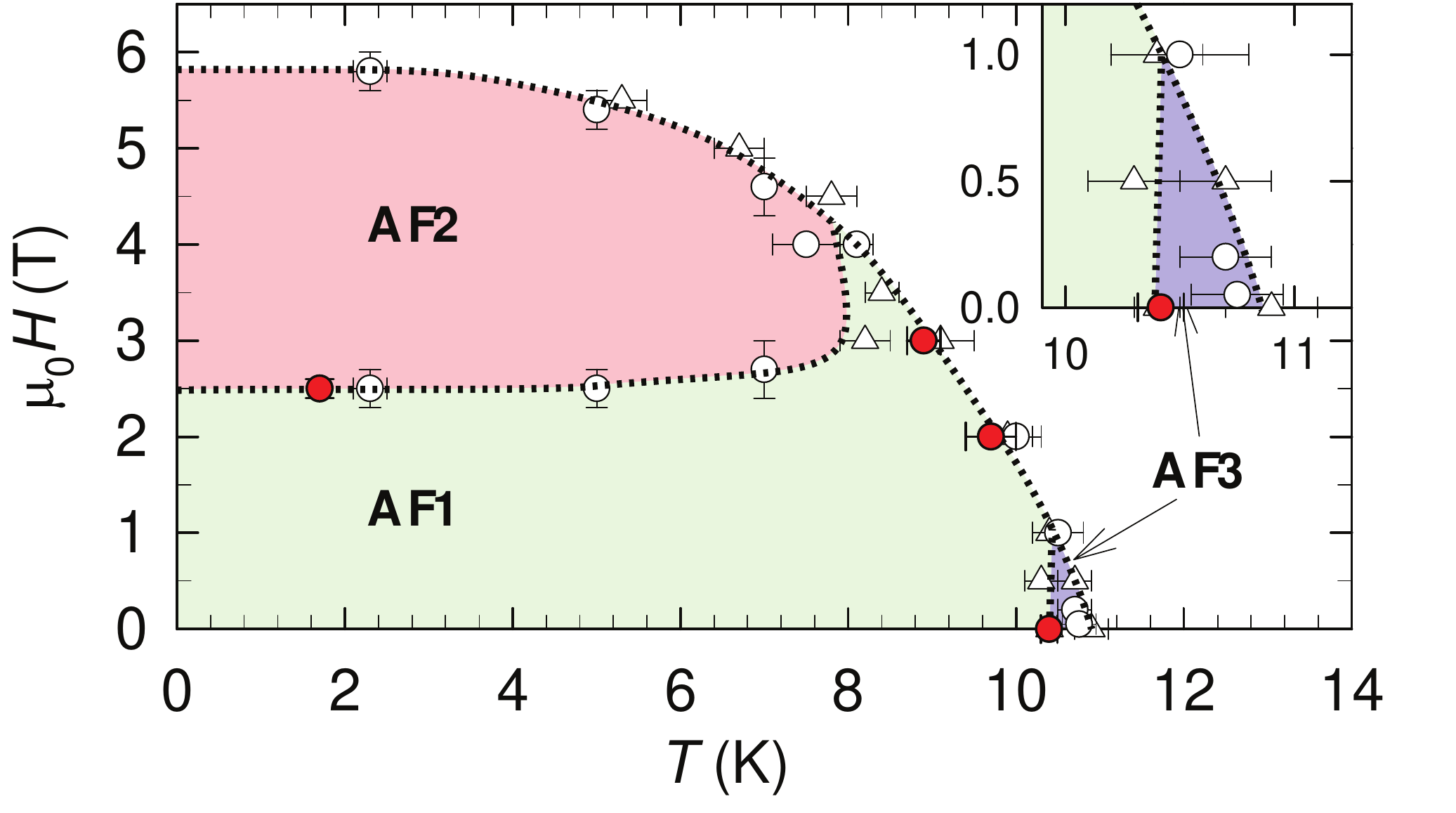}%
 }
 \caption{\label{phase-diagram} Magnetic phase diagram of Ho$_2$RhIn$_8$. Open points are determined from bulk measurements \cite{218-phase-diagram}, filled ones are from current neutron diffraction studies.
 Lines and shapes are to guide the eye.}
\end{figure}

Ho$_2$RhIn$_8$ orders antiferromagnetically below $T_N$ = 10.9~K and its magnetic properties
are driven mainly by crystal field effects (CEF) and the RKKY interaction \cite{218-phase-diagram}.
When an external magnetic field is applied along the tetragonal $c$ axis,
Ho$_2$RhIn$_8$ undergoes two successive magnetic phase transitions at 2~K.
Above 2.5~T an intermediate magnetic phase is established
and further increase of magnetic field up to 6~T results in the ferromagnetic order.
It is interesting to note, that the magnetization per formula unit in the AF2 phase is just a half of the moment found in
the ferromagnetic phase as was shown by magnetization measurements \cite{218-phase-diagram}.
The magnetic phase diagram of Ho$_2$RhIn$_8$ (see~Ref.~\onlinecite{218-phase-diagram} and Fig.~\ref{phase-diagram})
is qualitatively comparable to many other related "115" and "218" compounds with $R$ = Nd, Tb, Dy and Ho
where the $c$ axis is the easy magnetization axis:
$R$RhIn$_5$ \cite{RRh115-CF2007}, $R$CoIn$_5$ \cite{RCoGa5-Hudis}, $R$CoGa$_5$ \cite{RCoGa5-Hudis}, $R_2$RhIn$_8$ \cite{NdTb218,218-phase-diagram} and $R_2$CoGa$_8$ \cite{RCoGa218-bulk}.
As well as for Ho$_2$RhIn$_8$, the magnetization per formula unit in the field-induced AF2 phase in all these compounds is a half of the value in the ferromagnetic phase.
The microscopic nature of the magnetic order in the AF1 phase was studied in $R$RhIn$_5$ ($R$=Nd,Tb,Dy,Ho) \cite{RRh115-thesis}, $R$CoGa$_5$ ($R$=Tb,Ho) \cite{HoCoGa115-mag}, $R_2$RhIn$_8$ ($R$=Nd,Tb,Dy) \cite{218-magstruct} and $R_2$CoGa$_8$ ($R$=Tb,Dy,Ho) \cite{GdTbDyCoGa218,HoCoGa218},
resulting in the commensurate propagation $\textbf{k}_{115}$ = (1/2, 0, 1/2) in "115" compounds and $\textbf{k}_{218}$ = (1/2, 1/2, 1/2) in "218" compounds.
Magnetic structure in the AF2 phase was studied only by Hieu in his thesis \cite{RRh115-thesis} in NdRhIn$_5$ and DyRhIn$_5$,
however no conclusions from this measurements were given and magnetic structure in the AF2 phase remains unknown.

Despite sharing many similarities with the isostructural Nd, Tb and Dy compounds,
Ho$_2$RhIn$_8$ shows a feature not observed in other related compounds:
an additional magnetic phase existing in a narrow temperature region between $T_N$ = 10.9~K and $T_1$ = 10.4~K.
The jump in specific heat at $T_1$ is clearly pronounced and it was speculated,
that this phase transition is connected with a formation of an incommensurate magnetic phase \cite{218-phase-diagram}.

The present paper focuses on the determination of magnetic structures in all three magnetic phases of 
Ho$_2$RhIn$_8$ using several neutron diffraction experiments. Considering the similar phase diagram 
and results of magnetization measurements, the magnetic structure in the AF2 phase might be then 
eventually generalized for the whole class of related 115 and 218 tetragonal compounds.

\section{Experimental details}
The identical sample of Ho$_2$RhIn$_8$ as in the previous bulk experiments \cite{218-phase-diagram} was used.
The single crystal has the form of a cuboid of approximately 3 x 1 x 0.2 mm size.
No deviation from the known crystal structure (space group \emph{P4/mmm}, No.~123) nor any
presence of foreign phase were detected by X-Ray diffractometer and EDX microprobe analysis.
Magnetic susceptibility measurements were carried out with a physical property measurement
system (PPMS-14) by Quantum Design with the Vibrating Sample Magnetometer option 
at Magnetism and Low Temperatures Laboratories in Prague. 

Diffraction measurements were carried out on the several neutron diffractometers.
First, the nuclear structure and the extinction parameters were determined by
the D10 instrument at Institute Laue-Langevin (ILL). We used a PG002 monochromator
with an incident wavelength $\lambda = 2.36$~$\textrm{\AA}$ and a graphite filter before the sample.
The orientation matrix and lattice parameters were refined on the basis of 19 strong nuclear reflections at $T$ = 2~K.

Determination of the propagation vectors in the AF1 and AF3 phases was done on Laue diffractometer CYCLOPS \cite{CYCLOPS} at ILL.
Measurement of 2 patterns with a different sample rotation at $T$ = 14~K and in the ordered state at $T$ = 1.6~K took 3 hours per pattern.
In addition, a series of 50 patterns was taken in the slow temperature sweep mode (0.1 K/min)
in order to determine temperature dependence of the magnetic Bragg peaks and the nature of the AF3 phase.
Refinement of the Laue data were done in the Esmeralda Laue Suite program \cite{ESMERALDA}.

The magnetic phase AF1, as well as the behavior in applied magnetic fields,
were measured using the two-axis neutron diffractometer E4 at Helmholtz-Zentrum Berlin, Germany.
A focusing monochromator with vertically bent PG crystals was used for the wavelength $\lambda$ = 2.432 \AA.
The scattered intensities were recorded using a 200x200~mm two-dimensional position-sensitive detector (PSD) 896 mm from the sample.
The experiment was performed using a He flow cryostat at the temperature range 1.6 - 15 K.
First, the sample was loaded into a horizontal-field magnet, HM-2, and aligned with its reciprocal ($h$, 0, $l$) plane
in the horizontal scattering plane of the instrument. The magnetic field was applied along the easy $c$ axis.
In order to extend the number of observable reflections, the sample was realigned and mounted to the vertical-field magnet, VM-2,
to have ($h$, $k$, 0) plane aligned in the scattering plane of the instrument.
The 10 degrees opening angle of the magnet allowed us to reach reflections with index ($h$, $k$, 0.5) on the PSD detector.

We used the triple axis spectrometer IN3 at ILL to measure zero-field temperature dependence of magnetic Bragg reflections.
The sample was mounted similarly to the E4 experimental setup: with the reciprocal ($h$, 0, $l$) plane in the scattering plane 
and the experiment was performed in the elastic condition at $\lambda = 2.36$~$\textrm{\AA}$.

\section{Results and Discussion}

The data from D10 were integrated using the program RACER \cite{RACER}. Integration of the E4 data was done by a Python script; 
first cutting out background detector data
to a rectangular shape around the observed reflection and then fitting a Gaussian profile along the $\omega$-scan.
This technique allows us to reduce the background and also to distinguish out-of-plane and in-plane reflections.
Data from IN3 were fitted with a Gaussian profile, as this instrument has only 1D detector. All intensities were corrected for the Lorentz factor.
The obtained raw data were reduced using the program DataRed \cite{Fullprof} and the FullProf program \cite{Fullprof} was used for the refinement of the structures.

The structural parameters of Ho$_2$RhIn$_8$ were determined on the basis of 68 independent reflections measured on D10; they are summarized in Table~\ref{parameters}.
A single extinction parameter $q$ \cite{zachariasen,fpnews042002} was used in the FullProf software for correction of extinction, resulting in $q$=9.4(1.8).
This value combines both primary and secondary extinction proposed by Zachariasen \cite{zachariasen} and is suitable for isotropic description of extinction effects.
Extinction has not a negligible effect on strong magnetic reflections; 
the highest reduction of the intensity was observed for the reflection (1/2 0 1) which was reduced by coefficient 0.69.
As other neutron measurements were carried out using the the same wavelenght and the identical single crystal,
this extinction parameter was used as a fixed value for all E4 and IN3 data refinements.

\begin{table}
\caption{Refined structural parameters of Ho$_2$RhIn$_8$ at $T$=1.6~K. Space group: \emph{P4/mmm}. $a$=4.5648(16) \AA, $c$=11.953(12) \AA. } \label{parameters}
 \begin{ruledtabular}
     \begin{tabular}{lllll}
     atom      & Wyckoff pos.   & x & y & z \\
    \hline
     Ho    & 2$g$  &  0   & 0   &  0.3098(3)  \\
     Rh    & 1$a$  &  0   & 0   &  0         \\
     In(1) & 2$e$  &  0   & 0.5 &  0.5       \\
     In(2) & 2$h$  &  0.5 & 0.5 &  0.3086(6)  \\
     In(3) & 4$i$  &  0   & 0.5 &  0.1245(4)  \\
    \end{tabular}
  \end{ruledtabular}
\end{table}

\subsection{\label{sec:AF1}Zero field commensurate structure AF1}

\begin{figure}
\resizebox{0.5\textwidth}{!}{%
 \includegraphics{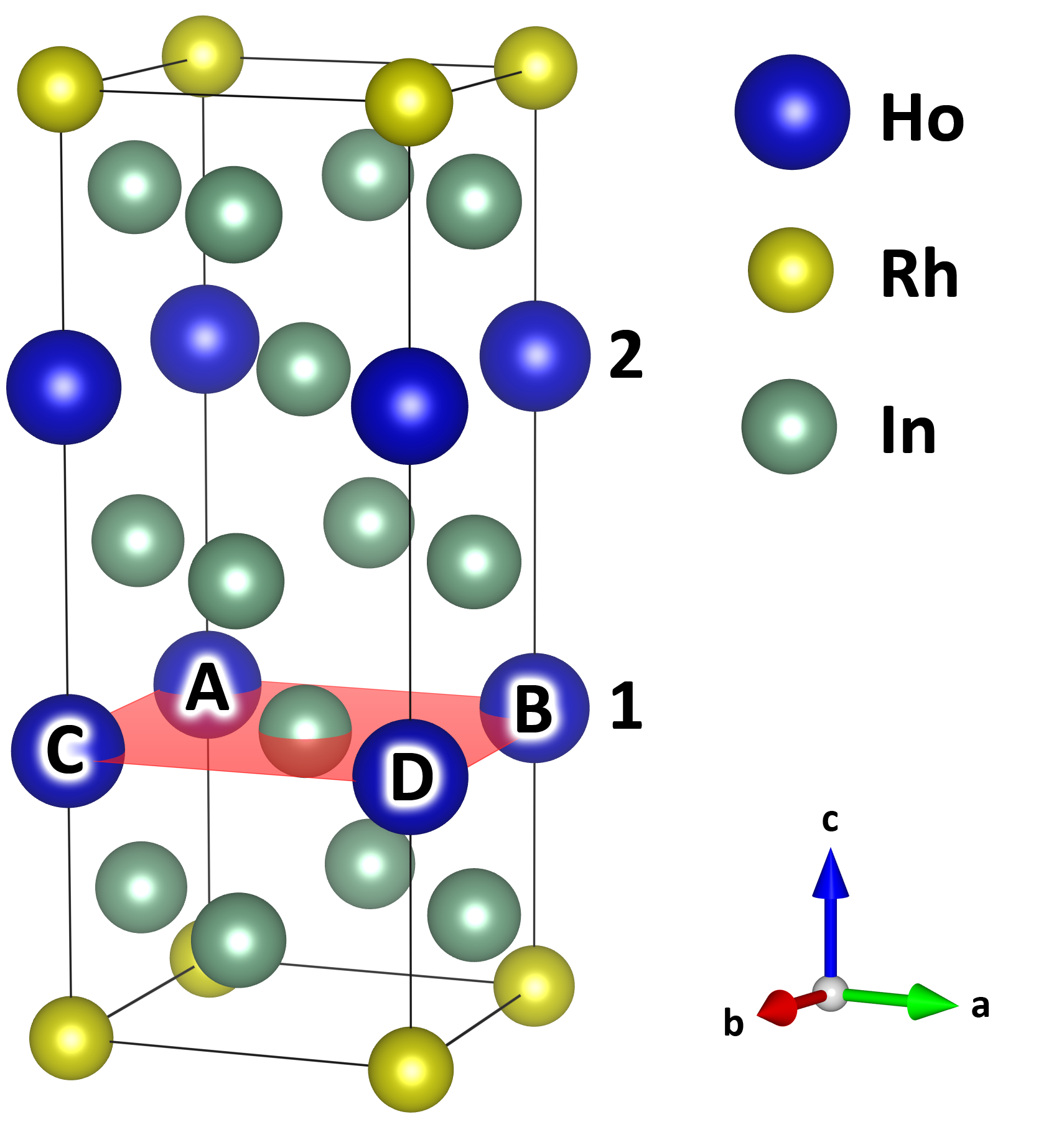}%
 }
 \caption{\label{crystal-structure} Crystal structure of the Ho$_2$RhIn$_8$ compound.}
\end{figure}

All observed diffraction spots on the total Laue pattern obtained in the paramagnetic state
can be indexed assuming a tetragonal structure with the space group \emph{P4/mmm}.
Their intensity remains unchanged when cooling to the ordered state at $T$ = 1.5 K,
showing no ferromagnetic or $\textbf{k}$ = (0, 0, 0) contribution to the magnetic structure in AF1.
A large number of purely magnetic reflections appears in the ordered state;
all of them can be described by a single propagation vector associated with the X [$\textbf{k}$ = (1/2, 0, 0) and $\textbf{k}'$ = (0, 1/2, 0)] point of symmetry.
There exist 8 1D irreducible representations associated with this wave vector, but only 6 of them are part of the global reducible magnetic representation of the 2$g$ Wyckoff site occupied by Ho atoms.
These can be assigned to 6 possible magnetic structures with the magnetic moments aligned along one of the main crystallographic directions,
each with ferro- or antiferromagnetic stacking of the moments on the two Ho positions within the unit cell (positions 1 and 2 in Fig.~\ref{crystal-structure}).
All possible structures are summarized in Table~\ref{table:posible-structures}.

Magnetic structure refinement using FullProf shows that magnetic moments lie along the $c$ axis, in agreement with magnetization data \cite{218-phase-diagram}.
The main difference between $++$ and $+-$ stacking along the $c$ axis is the existence of the (hk0) magnetic reflections.
In case of $+-$ stacking, all these reflections are forbidden. 
Indeed, we observe a zero magnetic intensity of (hk0) reflections favoring unambiguously the $+-$ stacking in Ho$_2$RhIn$_8$.
Detailed refinement of the intensities of the 7 reflections measured on the E4 diffractometer provides the size of the magnetic moment
$\mu_{AF1}$ = 6.9(2) $\mu_B$. Reliability factors of the fit are stated in Table \ref{table:reliability} in Appendix \ref{sec:relfactors}. 

\begin{table}
\caption{Possible magnetic structures in Ho$_2$RhIn$_8$ according to representations theory for given propagation vectors.
Moment directions are along the axes stated in the second column,
the stacking of the magnetic moments is over 2 unit cells along the $c$ axis on the positions 1 and 2 as shown in Fig.~\ref{crystal-structure}.} \label{table:posible-structures}
\begin{ruledtabular}
    \begin{tabular}{p{0.3\columnwidth}p{0.3\columnwidth}p{0.3\columnwidth}}
        prop.\ vector \newline \textbf{k} & moment \newline direction & $c$ axis \newline stacking  \\
    \hline
 (0, 0, 0) & $c$ & $++++$ \\
\hline
 \multirow{6}{*}{(1/2, 0, 0)} & $a$ & $+-+-$ \\
 & $a$ & $++++$ \\
 & $b$ & $+-+-$ \\
 & $b$ & $++++$ \\
 & $c$ & $+-+-$ \\
 & $c$ & $++++$ \\
\hline
 \multirow{6}{*}{(1/2, 0, 1/2)} & $a$ & $+--+$ \\
 & $a$ & $++--$ \\
 & $b$ & $+--+$ \\
 & $b$ & $++--$ \\
 & $c$ & $+--+$ \\
 & $c$ & $++--$ \\
\hline
 \multirow{4}{*}{(1/2, 1/2, 1/2)} & $c$ & $+--+$ \\
 & $c$ & $++--$ \\
 & in $ab$ plane & $+--+$ \\
 & in $ab$ plane & $++--$ \\
    \end{tabular}
 \end{ruledtabular}
\end{table}

\begin{figure}
\resizebox{0.5\textwidth}{!}{%
 \includegraphics{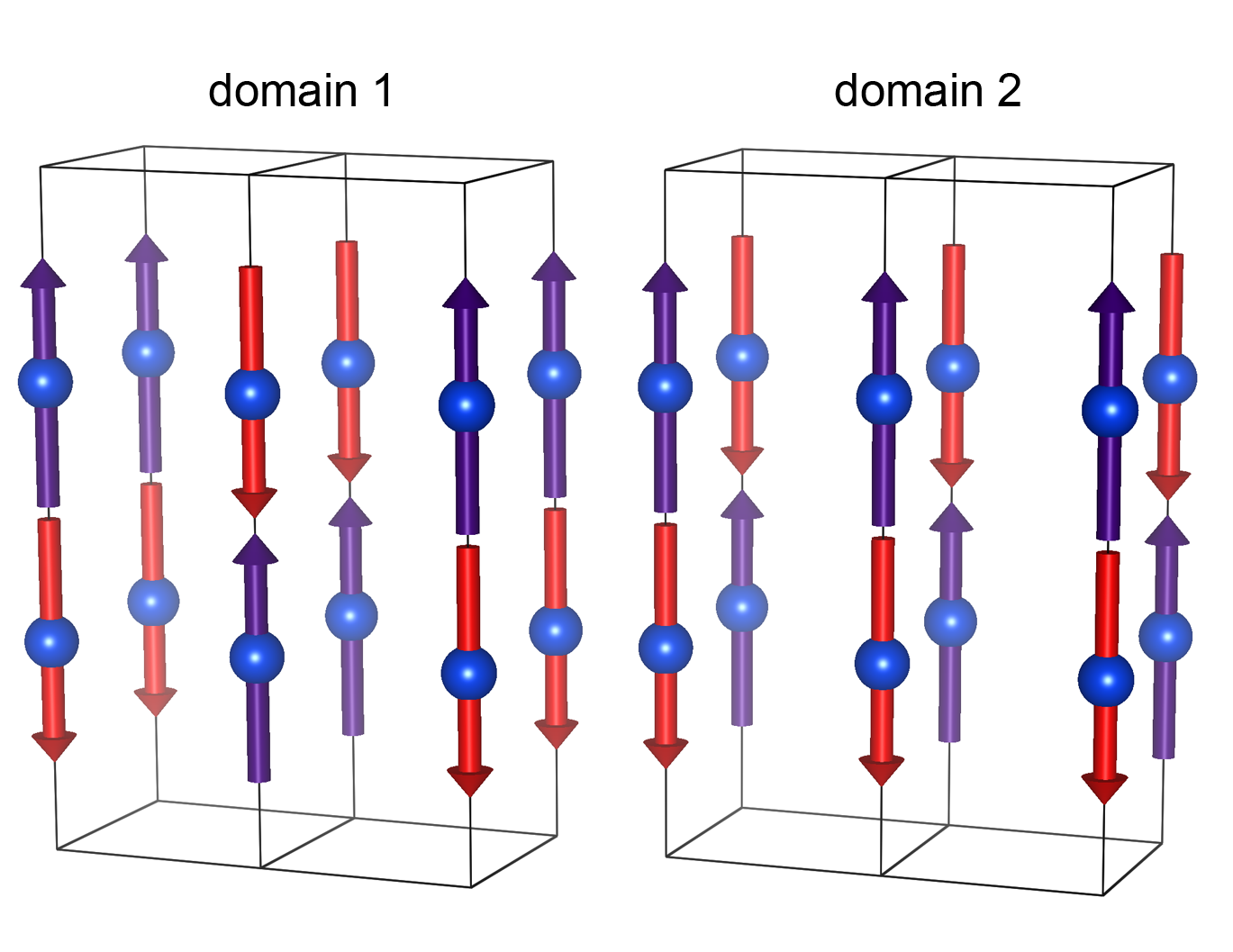}%
 }
 \caption{\label{struct-AF1} Magnetic structure of Ho$_2$RhIn$_8$ in the AF1 phase. Two magnetic domains are shown. Orientation is the same as in Fig.~\ref{crystal-structure}. }
\end{figure}

We have observed equally-sized magnetic Bragg reflections associated with both $\textbf{k}$ = (1/2, 0, 0) and $\textbf{k}'$ = (0, 1/2, 0) propagation vectors.
On the basis of a neutron diffraction experiment it is not possible to distinguish between multi-k structure and the existence of magnetic domains.
A multi-k structure would either require the moments lying out of the easy $c$ axis or it would imply the presence of nonmagnetic holmium atoms.
Both cases are unlikely with respect to magnetization measurements and therefore we conclude that there exist two magnetic k-domains,
corresponding to the propagation vectors $\textbf{k}$ = (1/2, 0, 0) and $\textbf{k}'$ = (0, 1/2, 0).
These domains are equally populated. Resulting magnetic structure is depicted in Fig.~\ref{struct-AF1}.

\begin{figure}
\resizebox{0.5\textwidth}{!}{%
 \includegraphics{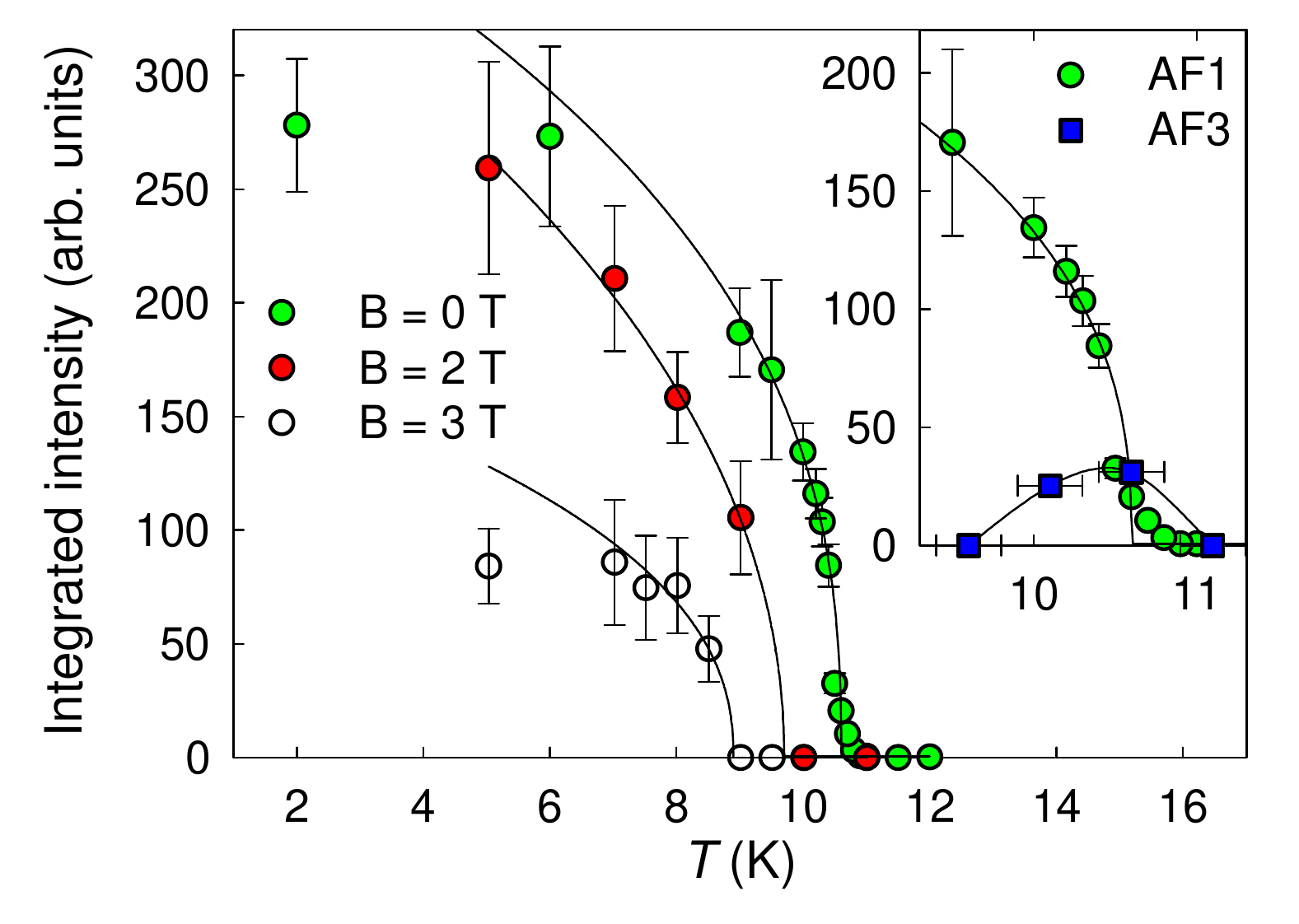}%
 }
 \caption{\label{Temp-dependence} Temperature dependence of integrated intensity of (1/2 0 1) reflection in different magnetic fields.
 Inset shows the intensity of the (1/2 0 1) reflection attributed to the AF1 phase (circles) together with (1/2 0.036 1) 
 reflection attributed to the AF3 phase (squares).
 Lines are only to guide the eye.}
\end{figure}

Fig.~\ref{Temp-dependence} shows the temperature dependence of the (1/2 0 1) magnetic reflection.
Transition temperature $T_1$ = 10.4(2) K was determined from the inflection point of this dependence.
It is in agreement with the value determined from specific heat \cite{218-phase-diagram} and is plotted in the phase diagram in Fig.~\ref{phase-diagram}.

Because the other members of the "218" compounds reveal $\textbf{k}_{218}$ = (1/2, 1/2, 1/2) \cite{218-magstruct},
Ho$_2$RhIn$_8$ is the first member of this series with magnetic domains.
Its structure is much more similar to the "115" compounds, revealing the identical stacking along the $c$ axis and also the same propagation within the $ab$ plane.

\subsection{Field induced structure AF2}

In order to determine magnetic structure in the field-induced magnetic phase AF2,
neutron diffraction experiments were carried out in the horizontal and vertical magnet in a field of 4~T.
A thorough search in the reciprocal space leads to observation of magnetic reflections described by 6 propagation vectors:
 $\textbf{k}_{0}$  = (0, 0, 0),
 $\textbf{k}_{1}$  = (1/2, 0, 0),
 $\textbf{k}_{1}'$ = (0, 1/2, 0),
 $\textbf{k}_{2}$  = (1/2, 0, 1/2),
 $\textbf{k}_{2}'$ = (0, 1/2, 1/2) and
 $\textbf{k}_{3}$  = (1/2, 1/2, 1/2), where $\textbf{k}_{1,2}$ and  $\textbf{k}_{1,2}'$ correspond to different magnetic domains.
Reciprocal space positions (1/2, 1/2, 0),  (1/2, 3/2, 0), (1, 1, 1/2) and (1, 0, 1/2) 
were measured for a longer time+ however, no magnetic reflections were found. 
The magnetic unit cell size is thus $2a$, $2b$, $2c$.
The sum of the magnetic moments associated with $\textbf{k}_{1}$, $\textbf{k}_{2}$ and $\textbf{k}_{3}$ 
within the magnetic unit cell is always zero, as they are always propagating within the $ab$ plane canceling out moments at the 2$g$ Wyckoff site.
The magnetization measurement clearly shows that the overall magnetic moment in the AF2 phase amounts to the half of the magnetic moment of Ho in ferromagnetic state above 6~T
($\approx$ 4 $\mu_B$)\cite{218-phase-diagram}. This moment does not propagate and is associated with the ferromagnetic component ($\textbf{k}_{0}$) of the AF2 phase.
Refinement of the measured ferromagnetic intensities using the FullProf software confirmes this prediction
and leads to the magnetic moment $\mu_{\mathbf{k}_{0}}$ = 3.6(4) $\mu_B$ in agreement with magnetization measurements.

We have performed symmetry analysis based on representations theory for the remaining propagation wave vectors similarly to AF1.
Possible directions of the magnetic moments are summarized in Table~\ref{table:posible-structures}.
For structures with $\textbf{k}_{1}$ and $\textbf{k}_{2}$ there exist 6 allowed 1D irreducible representations.
In case of $\textbf{k}_{3}$, there is a general possibility for the moment to lie in any direction within the $ab$ plane
(detailed symmetry and magnetic group analysis for this structure and propagation is discussed in Ref. \onlinecite{218-magstruct}).
Taking into account the fact, that all moments in AF1 point along the $c$ axis and that a clear field-induced spin-flip behavior is observed,
we consider that magnetic moments associated with every wave vector $\textbf{k}$ in AF2 point also along the $c$ axis.

In order to construct possible magnetic structures, we will now focus on the magnetic spin arrangement 
within the (00$z_\textrm{Ho}$) plane (highlighted in red color in Fig.~\ref{crystal-structure}).
This consideration simplifies the problem to 2 dimensions.
There are 4 magnetic positions within this plane in the magnetic unit cell (marked A-D in Fig.~\ref{crystal-structure}), 
all corresponding to one atom site in the nuclear unit cell.
There can exist 4 propagation vectors in maximum:
 $\textbf{k}_{0,plane}$  = (0, 0),
 $\textbf{k}_{1,plane}$  = (1/2, 0),
 $\textbf{k}_{2,plane}$  = (0, 1/2) and
 $\textbf{k}_{3,plane}$  = (1/2, 1/2).
Considering only spin-flip scenario, the total magnetic moments on all 4 sites $\mu_{A-D}$ have necessarily the same amplitude.
If we neglect the change of the magnitude of the magnetic moment and assume that the magnetic moments are along the $c$ axis,
the only possible solution exists:
\begin{equation}
{\mu_{\mathbf{k}_0} = \mu_{\mathbf{k}_1} = -\mu_{\mathbf{k}_{2}} = \mu_{\mathbf{k}_{3}}},\label{eq:propvectorrelation}
\end{equation}
\begin{equation}
{\mu_{A} = -\mu_{B} = \mu_{C} = \mu_{D} = 2\mu_{\mathbf{k}_0}}.\label{eq:sitemomentrelation}
\end{equation}
In other words, the component associated with $\mathbf{k}_{2}$ has an opposite sign with respect to all other components and one of the 4 magnetic moments within the plane is flipped.
For details of the derivation, see Appendix \ref{sec:magnetic_details}.

\begin{figure}
\resizebox{0.5\textwidth}{!}{%
 \includegraphics{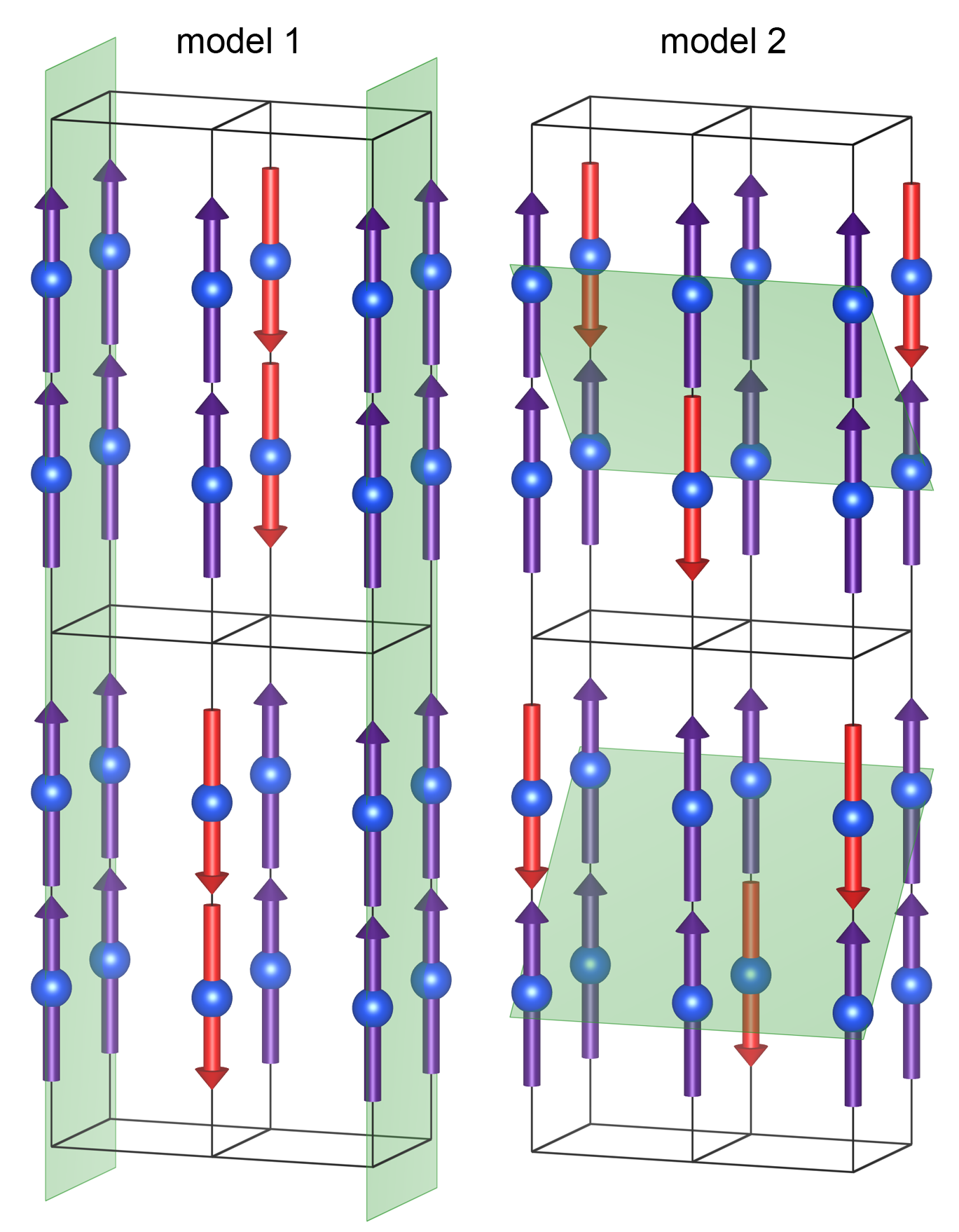}%
 }
 \caption{\label{AF2-models} {Possible magnetic structures in the AF2 phase with highlighted ferromagnetically ordered planes.
 Model 1 consists of ferromagnetic and antiferromagnetic (00l) planes.
 Model 2 is more complicated revealing ferromagnetically ordered atoms in some kind of zigzag planes.
 Orientation is the same as in Fig.~\ref{crystal-structure}.}}
\end{figure}

\begin{figure}[b]
\resizebox{0.5\textwidth}{!}{%
 \includegraphics{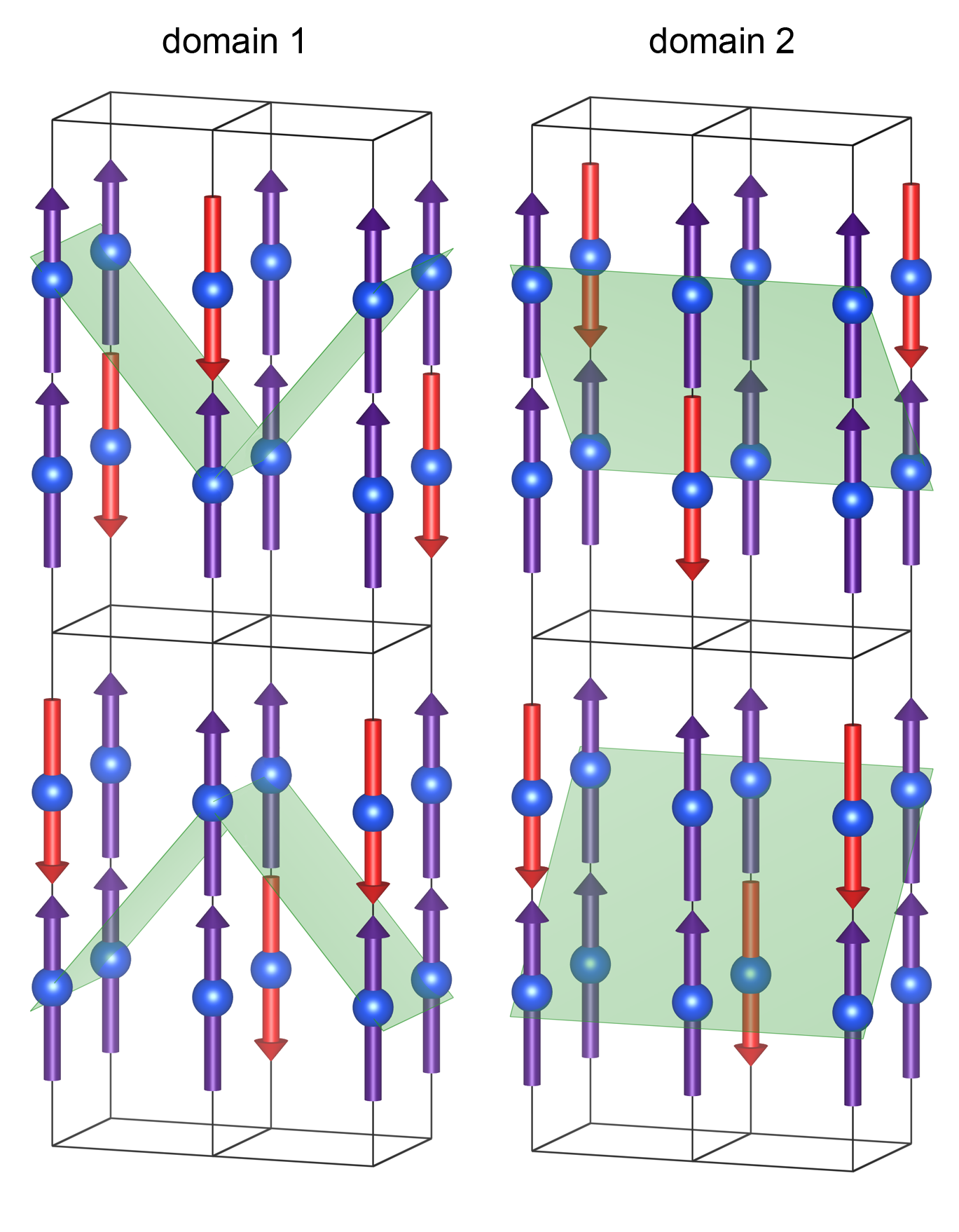}%
 }
 \caption{\label{struct-AF2} The magnetic structure of Ho$_2$RhIn$_8$ in the AF2 phase. Two magnetic domains are equally populated. The orientation is the same as in Fig.~\ref{crystal-structure}. }
\end{figure}

Extending from 2D case to the real Ho$_2$RhIn$_8$ structure brings more options 
(6 propagation vectors, each with 2 possible stackings along the $c$ axis).
Taking into account the fact, that the total magnetic moment at one site cannot be bigger than full magnetic moment of Ho (10~$\mu_B$) and $\mu_{\mathbf{k}_{0}} = 3.6(4) \mu_B$,
only two independent models summarized in Table~\ref{table:AF2-models} and depicted in Fig.~\ref{AF2-models} are possible.
They are distinguishable on the same principle as in the case of the AF1 phase - on
the basis of the existence of the reflections (hk0) for propagation vector $\textbf{k}_{1}$.
These reflections are forbidden in the model 2. As we indeed did not observe any of these reflections,
the correct model describing the magnetic structure of Ho$_2$RhIn$_8$ in the AF2 phase is the model 2.
Similarly to AF1, there exist two magnetic k-domains. The resulting magnetic structure is shown in Fig.~\ref{struct-AF2}.

\begin{table}
\caption{Possible magnetic structures in the AF2 phase based on the propagation vectors analysis. 
 Stacking along the $c$ axis has the same meaning as in Table~\ref{table:posible-structures}.} \label{table:AF2-models}
 \begin{ruledtabular}
    \begin{tabular}{lll}
        & \multicolumn{2}{c}{$c$ axis stacking on site A (see Fig.~\ref{crystal-structure})} \\
        &	model 1 & model 2 \\
    \hline
 $\textbf{k}_{0}$ = (0,     0,   0) & $++++$ & $++++$ \\
 $\textbf{k}_{1}$ = (1/2,   0,   0) & $++++$ & $+-+-$ \\
 $\textbf{k}_{2}$ = (0,   1/2, 1/2) & $--++$ & $+--+$ \\
 $\textbf{k}_{3}$ = (1/2, 1/2, 1/2) & $++--$ & $++--$ \\
    \hline
 Overall stacking                   & $++++$ & $+++-$ \\
    \end{tabular}
 \end{ruledtabular}
\end{table}

Quantitative refinement using the FullProf software confirmed described results and led to the magnetic moments
associated with different propagation vectors as follows: $\mu_{\mathbf{k}_{0}}$ = 3.6(4) $\mu_B$,
$\mu_{\mathbf{k}_{1}}$ = 3.7(3) $\mu_B$ and $\mu_{\mathbf{k}_{2}}$ = -4.0(2) $\mu_B$.                  
The amplitude of the magnetic moments described by the propagation vector $\textbf{k}_{3}$ was not possible to determine,
since due to the construction of magnets, we have reached only one magnetic reflection associated with this propagation.
Reliability factors of the fitting procedure are stated in Table \ref{table:reliability} in Appendix \ref{sec:relfactors}.
All 3 determined amplitudes satisfy equation (\ref{eq:propvectorrelation}) within their errors.
The overall amplitude of magnetic moments is $\mu_{AF2}$ = 7.5(5) $\mu_B$,
which is calculated from equation (\ref{eq:sitemomentrelation}) taking $\mu_{\mathbf{k}_0}$ as the mean of all 3 refined amplitudes of magnetic moments.
The value of $\mu_{AF2}$ is slightly bigger than $\mu_{AF1}$. This increase is due to the impact of the 4~T external magnetic field and is in agreement with the measured magnetization curves \cite{218-phase-diagram}.

\begin{figure}
\resizebox{0.5\textwidth}{!}{%
 \includegraphics{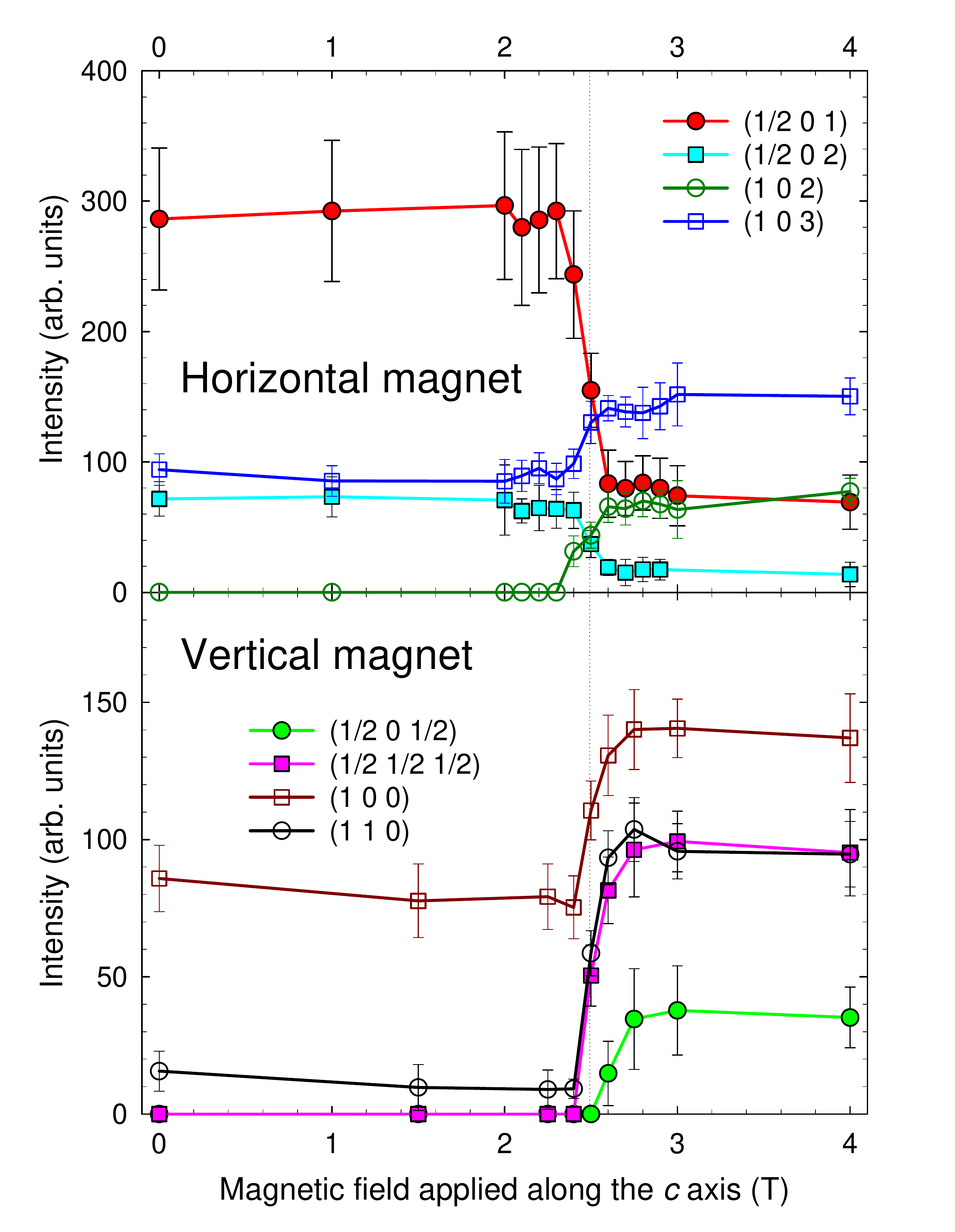}%
 }
 \caption{\label{Field-dependence} Field dependence of selected Bragg reflections in horizontal and vertical magnetic fields. Lines are there only to guide the eye.}
\end{figure}

To clarify the location of the phase boundaries and verify consistency of the data from vertical and horizontal magnet,
several reflections were followed with the changing magnetic field (Fig.~\ref{Field-dependence}).
The transition from the AF1 to the AF2 phase is illustrated by a strong decrease of intensity of the (1/2 0 1) and (1/2 0 2) reflections
together with increase of the intensity at the positions of the nuclear peaks and peaks described by the propagation vectors $\textbf{k}_{2}$ and $\textbf{k}_{3}$.
Temperature dependence of the (1/2 0 1) reflection in the fields of 2 and 3 T is depicted in Fig.~\ref{Temp-dependence}. The shape of the curve in 0 T and 2 T corresponds to each other showing the same ordering mechanism as both are entering the AF1 phase.
The emergence of the AF1 phase within a limited temperature range in the field of 3 T, indicated by bulk measurements (Fig.~\ref{phase-diagram}), was not observed.
This can be explained by absence of a long-range order in the narrow AF1 phase region just below the ordering temperature in 3 T.
Points from the measured temperature and the field dependencies are included in the phase diagram in Fig.~\ref{phase-diagram}.

The AF2 phase in Ho$_2$RhIn$_8$ is the first solved field-induced magnetic structure within the $R_2T$In$_8$ and $RT$In$_5$ family
preventing us from comparisons with related compounds.
We suppose that, due to the similar phase diagrams,
related compounds from the "218" family reveal the same flipping mechanism during metamagnetic transition from AF1 to AF2,
consisting of the flip of 1/4 of the magnetic moments.
The same assumption could be applied with a small modification to the "115" compounds, 
as the AF1 structure in Ho$_2$RhIn$_8$ is very similar to the known "115" magnetic structures.
On the basis of the magnetization measurements, Hieu suggested several possible magnetic structures in AF2 phase for "115" compounds \cite{RRh115-thesis}.
One of these magnetic structures (Fig. 5.69(c) in Ref. \onlinecite{RRh115-thesis}) corresponds to the model determined for the AF2 phase in Ho$_2$RhIn$_8$.

\subsection{\label{sec:AF3}Incommensurate structure AF3}

\begin{figure}
\resizebox{0.5\textwidth}{!}{%
 \includegraphics{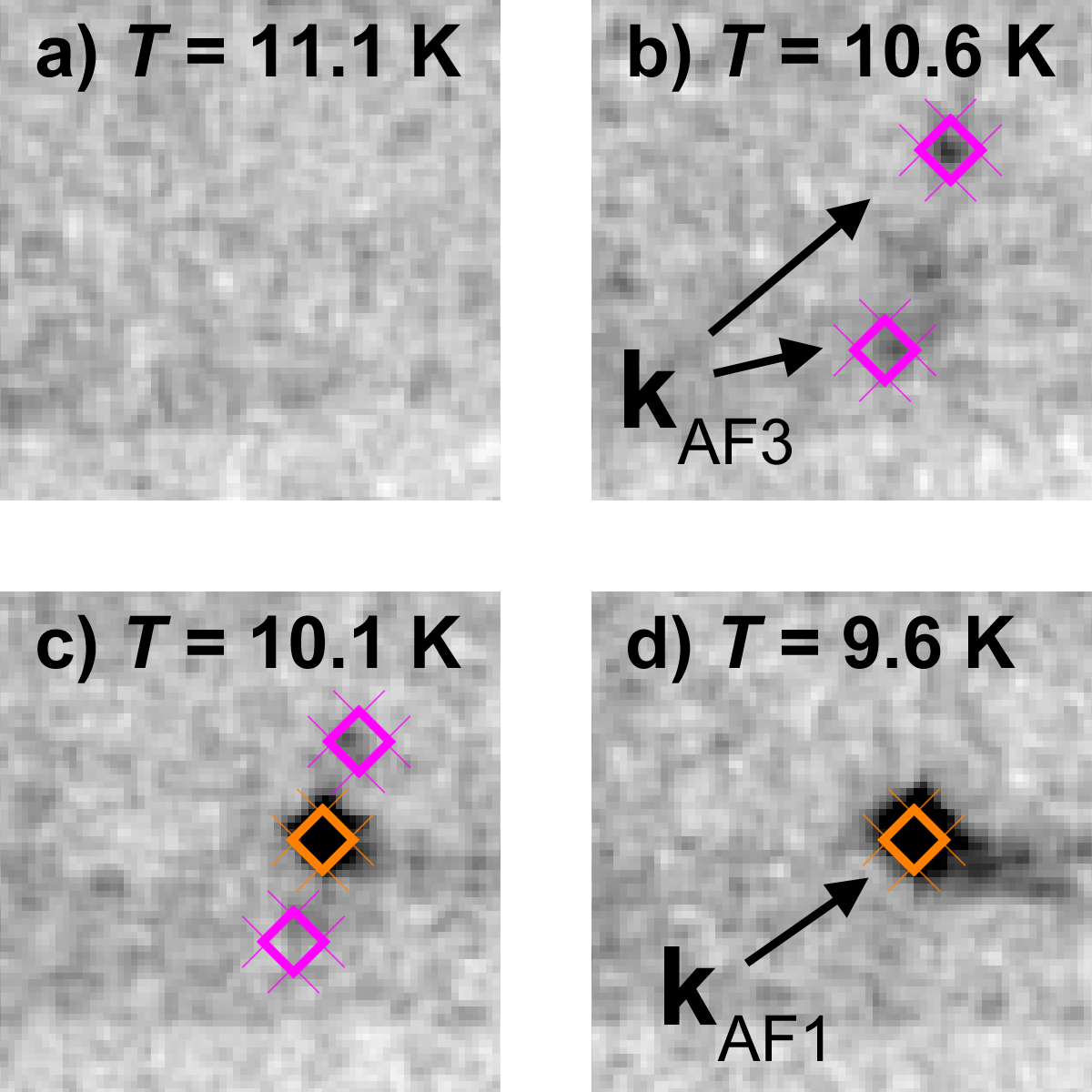}%
 }
 \caption{\label{Laue-zoom} Detail of the region in the vicinity of (0.5 0 1) reflection in the Laue pattern taken at different temperatures.}
\end{figure}

\begin{figure}
\resizebox{0.5\textwidth}{!}{%
 \includegraphics{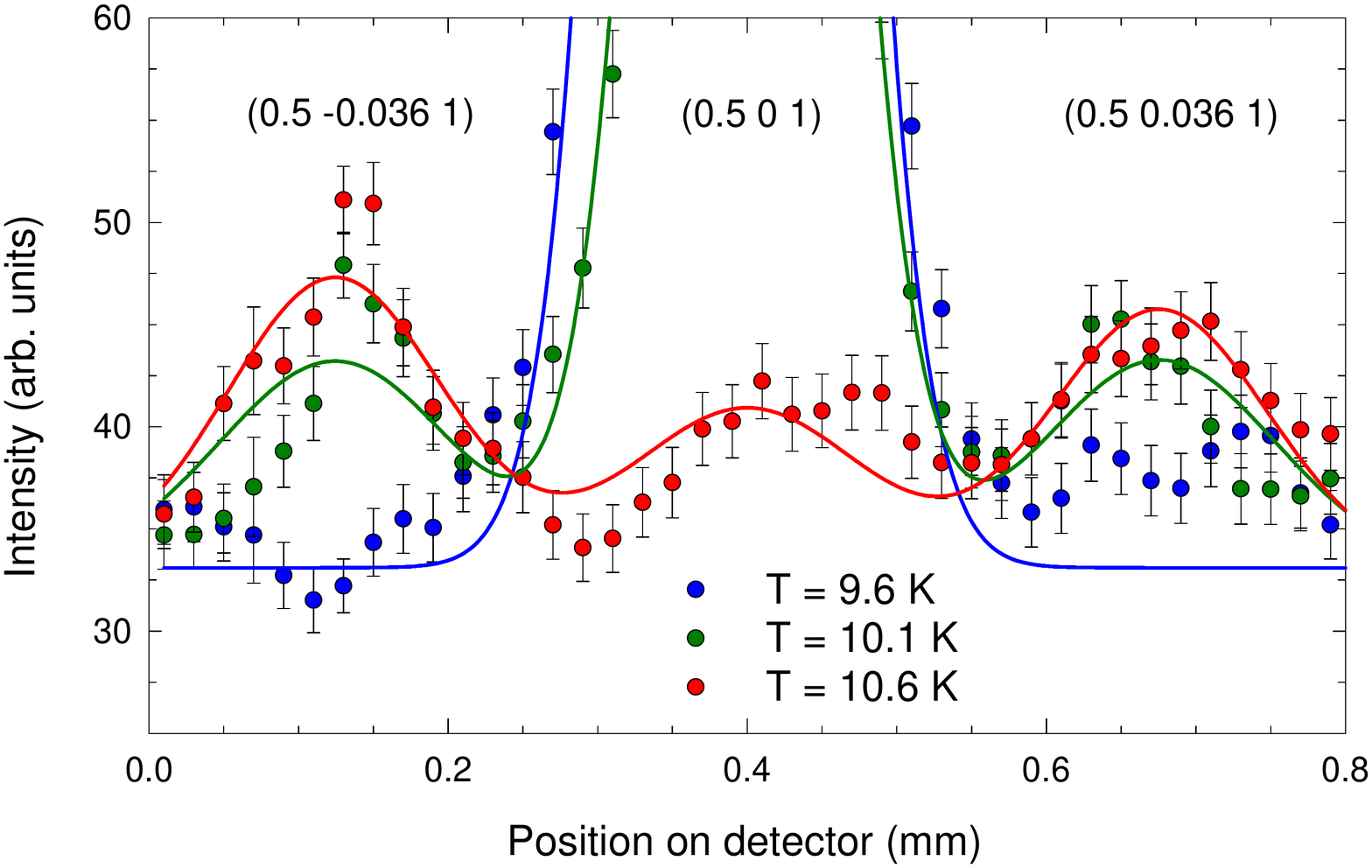}%
 }
 \caption{\label{Laue-cuts} Cuts along the 0h0 crystallographic direction at different temperatures.}
\end{figure}

Magnetic intensity in the AF3 phase is illustrated in Fig.~\ref{Laue-zoom},
which shows the identical detail of the Laue diagrams taken at different temperatures.
At 11.1 K, above the N\'eel temperature, there is no significant intensity.
At 10.6 K, two satellites at incommensurate position appear together with a very weak trace of a commensurate (1/2 0 1) reflection.
The intensity of the commensurate reflection starts to grow and at 10.1 K there are clearly both commensurate and incommensurate reflections visible.
The AF3 phase completely vanishes at 9.6~K.
Integrated cut along the curves going through all three reflections are shown in Fig.~\ref{Laue-cuts}.
The same behavior was observed also around other strong magnetic reflections and the magnetic peaks on the incommensurate positions were indexed with the propagation vector $\textbf{k}_{AF3}$ = (1/2, $\delta$, 0),
where $\delta$ = 0.036(3). However, as the Laue patterns were taken during temperature sweep and all spots are well localized in reciprocal space at constant positions,
$\delta$ is temperature independent. This can be also seen in Fig.~\ref{Laue-cuts}.
The data are consistent with $T_N$ = 10.9~K determined from specific heat measurements \cite{218-phase-diagram}.

Formation of the incommensurate zero-field phase is unique within "218" and "115" compounds,
but it can be found in the tetragonal compounds structurally related to the other well known heavy fermion superconductor CeCu$_2$Si$_2$ (e.g. in UCu$_2$Si$_2$ \cite{UCu2Si2}).
A tiny value of the incommensurate component of the propagation vector implicates modulation period involving about 27 holmium atoms.
Such a long modulation can be explained by formation of a spin-density-wave phase.

\subsection{\label{sec:CF}Crystalline electric field}

\begin{figure}
\resizebox{0.5\textwidth}{!}{%
 \includegraphics{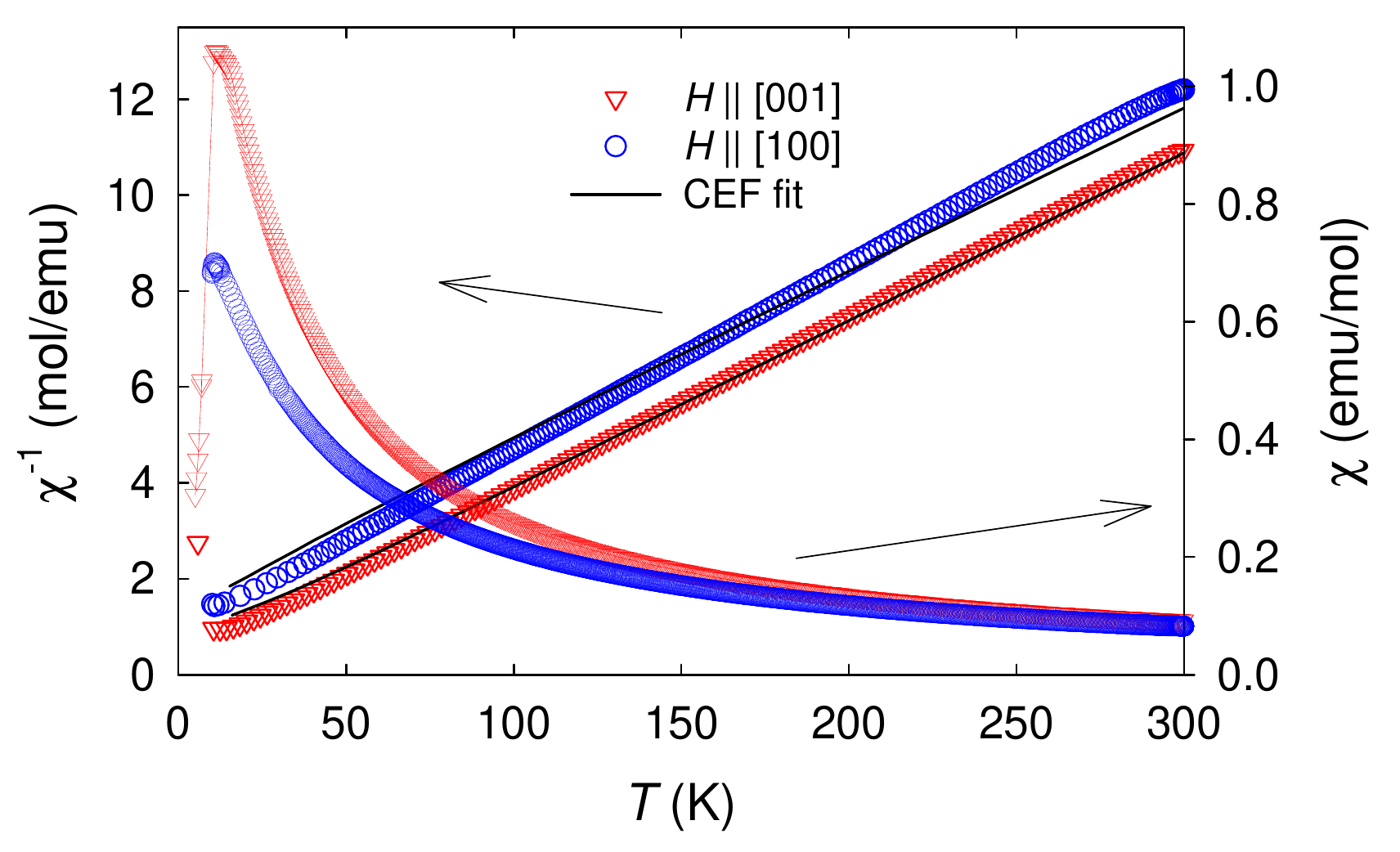}%
 }
 \caption{\label{Susceptibility} Temperature dependence of magnetic susceptibility 
 (right axis) and inverse magnetic susceptibility (left axis) of Ho$_2$RhIn$_8$. 
 Values are given per one mole, thus 2 Ho atoms. Solid line represents CEF fit of inverse susceptibility data.}
\end{figure}

In order to understand the anisotropy in  Ho$_2$RhIn$_8$, 
we have measured temperature dependence of the magnetic susceptibility 
in the applied magnetic field of 1~T along the crystallographic $a$ and $c$ axis. 
Inverse susceptibility data in the paramagnetic region (30 - 300K) were fitted to 
the modified Curie-Weiss law,
\begin{equation}
{\chi = \chi_0 + \frac{C}{T-\theta_p}},\label{eq:cwlaw}
\end{equation}
where $\chi_0$ is temperature independent susceptibility, 
$\theta_p$ is paramagnetic Curie temperature and the effective magnetic moment $\mu_{\rm eff}$ is related to Curie constant $C$.
Measured data are shown in Fig. \ref{Susceptibility} and 
determined parameters for each field direction are summarized in Table \ref{table:cfpars}. 
We observe expected magnetic anisotropy as difference between two measured field directions.
The rapid decrease of the magnetic 
susceptibility below ordering temperature in field along [001] direction
confirms antiferromagnetic ordering with Ho moments along the $c$ axis.

\begin{table*}[!ht]
 \caption{Summary of parameters obtained from the fit to Equation \ref{eq:cwlaw} and determined crystal field parameters (Eq. \ref{eq:susc}).} \label{table:cfpars}
 \begin{ruledtabular}
  \begin{tabular}{llllll}
      & $\chi_0$ &  $\theta_p$ & $\mu_{\rm eff}$ & $\lambda$ & $\mathcal{J}_{\rm ex}$  \\
      & (emu/mol)&  (K)        & ($\mu_B$)   & (mol/emu) & (K)       \\
    \hline
$[100]$ & $2.1(1)\times10^{-3}$ & -21.1(1)   & 10.08(2)  & -1.31     & -0.68     \\
$[001]$ & $0.5(1)\times10^{-3}$ & -8.9(2)    & 10.57(2)  & -0.58     & -0.78     \\
    \hline
      & $B_2^0$ (K) & $B_4^0$ (K)         & $B_4^4$ (K)          & $B_6^0$ (K)          & $B_6^4$ (K)         \\
      & -0.173      & $1.08\times10^{-3}$ & $-1.27\times10^{-2}$ & $-3.35\times10^{-6}$ & $9.70\times10^{-6}$  \\
  \end{tabular}
 \end{ruledtabular}
\end{table*}

Further analysis of susceptibility data was done by fitting to CEF model 
proposed by Stevens and Hutchings \cite{hutchingsCEF}. 
Details of this procedure as well as all used formulas are described in Appendix \ref{sec:cf-fit}.
Solid lines in Fig. \ref{Susceptibility} are the results of CEF calculations which reproduce well the measured data.
The CEF parameters and molecular field coefficients $\lambda$ are summarized in Table \ref{table:cfpars}.
Calculated second order parameter $B^0_2$ is clearly dominant, in agreement 
with related $R$RhIn$_5$ \cite{RRh115-CF2007} and $R_2$CoGa$_8$ \cite{RCoGa218-bulk} compounds.
It is reported for both series, that sign of $B^0_2$ parameter determines 
the easy magnetization axis. The change of the easy axis of magnetization takes place 
between Ho and Er compound within each series (discussed also in Ref. \onlinecite{RCoGa5-Hudis}).
This can be understood by change of the sign of Stevens multiplication factor $\alpha_J$ for rare-earth ions, because
$B^0_2$ is defined as $A^0_2 \langle r^2 \rangle \alpha_J$. 
Here $A^0_2$ is constant for identical structures and $r^2$ is always positive.  
As stated at the end of Section \ref{sec:AF1}, zero-field magnetic structure AF1 in Ho$_2$RhIn$_8$
differs from magnetic structures in other $R_2$RhIn$_8$ compounds. 
Moreover, it is also different from its cobalt-gallium analogue Ho$_2$CoGa$_8$
which has magnetic moments propagating with wave vector $\textbf{k}$ = (1/2, 1/2, 1/2) \cite{HoCoGa218}
in agreement with other rhodium-indium compounds. 
In fact, Ho$_2$RhIn$_8$ is the only compound from the rich $R_{n}T_{m}X_{3n+2m}$ family
with magnetic moments pointing along the $c$ axis and the propagation vector $\textbf{k}$ = (1/2, 0, 0).
Determined $B^0_2$ = -0.17~K for Ho$_2$RhIn$_8$ is smaller than values reported earlier for related 
HoRhIn$_5$ (-0.33~K \cite{RRh115-CF2007}) and Ho$_2$CoGa$_8$ (-0.22~K \cite{RCoGa218-bulk})
and is closer to the zero, where the easy $c$ axis switches to the easy $ab$ plane.
Competing $c$ and $ab$ easy axis/planes in Ho$_2$RhIn$_8$ can stand behind the
unique anisotropic propagation within the $ab$ plane and also existence \
of the zero-field incommensurate phase AF3.   

We have calculated exchange interaction 
$\mathcal{J}_{\rm ex}$ according to the mean field theory by using the same formulas
as in Ref. \onlinecite{RCoGa218-bulk}. The determined values are presented in Table \ref{table:cfpars}.
Both exchange constants are negative confirming antiferromagnetic interaction,
but they are significantly increased in comparison with Ho$_2$CoGa$_8$ \cite{RCoGa218-bulk}.
This effect is reflected in higher ordering temperature of Ho$_2$RhIn$_8$.

\section{Conclusion}

We have determined the magnetic structures in Ho$_2$RhIn$_8$ by means of neutron diffraction experiments.
In the zero-field ordered state, the magnetic order is characterized by a single propagation vector $\textbf{k}_{AF1}$ = (1/2, 0, 0)
with the antiferromagnetic coupling of Ho moments along the $c$ axis and the amplitude of the magnetic moment 6.9(2)~$\mu_B$.
Before entering this ground-state commensurate phase, there exists a small temperature region 
where incommensurate phase with propagation $\textbf{k}_{AF3}$ = (1/2, 0.036, 0) develops.
Both existence of the additional incommensurate phase and the ground state propagation vector anisotropic within $ab$ plane 
are unique within the $R_{n}T_{m}X_{3n+2m}$ series.
Analysis of the temperature dependence of magnetic susceptibility revealed 
paramagnetic Curie temperatures and set of crystal field parameters
which are in agreement with observed anisotropic magnetic structures.
These results confirm that magnetism in Ho$_2$RhIn$_8$ is purely CEF driven. Mean field theory
model can be used for the estimation of the magnetic exchange constants and magnetic structures. 

In the applied magnetic field along the $c$ axis magnetic structure transforms
by spin-flipping of the 1/4 of magnetic moments to another commensurate phase with the amplitude of magnetic moment 7.5(5)~$\mu_B$ (in $B$ = 4~T).
Similar flipping behavior is expected also in the other related compounds with the same phase diagram.

\begin{acknowledgments}
This work was supported by the Czech Science Foundation under Grant No. P204-13-12227S.
The work is a part of research project LG14037 financed by the Ministry of Education, Youth and Sports, Czech Republic.
Sample preparation and bulk measurements were carried out in MLTL (http://mltl.eu/), which is supported within the program of Czech Research Infrastructures (project No. LM2011025).
This project has received funding from the European Union's Seventh Framework Programme for research, technological development and demonstration under the NMI3-II Grant number 283883.
\end{acknowledgments}

\appendix
\section{\label{sec:magnetic_details}Magnetic structure derivation}

Generally, moment distribution $\mu_j$ associated with the atom $j$ of a magnetic structure can be Fourier expanded according to:
\begin{equation}
{\displaystyle \mu_{j}=\sum_{\left\{ \mathbf{k}\right\} } \Psi^{\mathbf{k}}_{j}\mathrm{e}^{-2\pi i \mathbf{k\cdot}\mathbf{T}}},\label{eq:propvectordefn}
\end{equation}
where $\Psi^{\mathbf{k}}_{j}$ is the basis vector associated with the propagation vector $\mathbf{k}$ and the atom on the position $j$
and $\mathbf{T}$ is the corresponding lattice translation vector.
The summation is made over all wave vectors that are confined to the first Brillouin zone.
If we neglect the change of the magnitude of the magnetic moment and assume 
that the magnetic moments lie along the $c$ axis,
equation (\ref{eq:propvectordefn}) will simplify to:
\begin{equation}
{\displaystyle \mu_{j}=\sum_{\left\{ \mathbf{k}\right\} }\mu_{\mathbf{k}}\cos(\mathbf{k\cdot}\mathbf{T})},\label{eq:simplified}
\end{equation}
where cosine has always maximal amplitude.

\begin{figure}[t]
\resizebox{0.5\textwidth}{!}{%
 \includegraphics{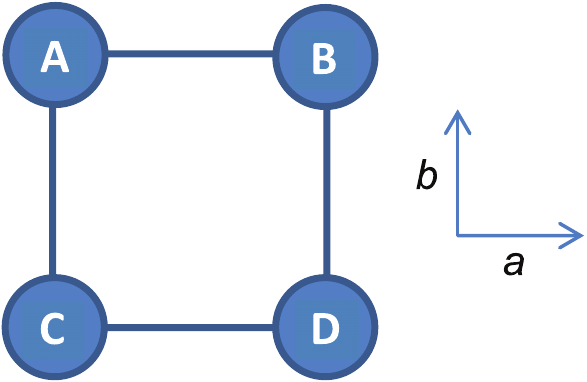}%
 }
 \caption{\label{diagram} Schematic view on Ho atoms in $ab$ plane of magnetic unit cell. Each atom position is marked with a letter.}
\end{figure}

We will now focus on the magnetic spin arrangement within $2a$ x $2b$ plane, 
where are four magnetic positions in the magnetic unit cell (marked A-D in Fig.~\ref{diagram}), all corresponding to one atom site in the nuclear unit cell.
There exist 4 propagation vectors in maximum:
 $\textbf{k}_{0,plane}$  = (0, 0),
 $\textbf{k}_{1,plane}$  = (1/2, 0),
 $\textbf{k}_{2,plane}$  = (0, 1/2) and
 $\textbf{k}_{3,plane}$  = (1/2, 1/2). In the following text, subscript $plane$ will be omitted for better readability. Regarding equation (\ref{eq:simplified}), total magnetic moments on A - D sites can be calculated as:
\begin{equation}
{\displaystyle \mu_A=\mu_{\mathbf{k}_0}+\mu_{\mathbf{k}_1}+\mu_{\mathbf{k}_2}+\mu_{\mathbf{k}_3}},\label{eq:propA}
\end{equation} 
\begin{equation}
{\displaystyle \mu_B=\mu_{\mathbf{k}_0}-\mu_{\mathbf{k}_1}+\mu_{\mathbf{k}_2}-\mu_{\mathbf{k}_3}},\label{eq:propB}
\end{equation} 
\begin{equation}
{\displaystyle \mu_C=\mu_{\mathbf{k}_0}+\mu_{\mathbf{k}_1}-\mu_{\mathbf{k}_2}-\mu_{\mathbf{k}_3}},\label{eq:propC}
\end{equation} 
\begin{equation}
{\displaystyle \mu_D=\mu_{\mathbf{k}_0}-\mu_{\mathbf{k}_1}-\mu_{\mathbf{k}_2}+\mu_{\mathbf{k}_3}},\label{eq:propD}
\end{equation} 

Considering the spin-flip ratio only, the total magnetic moments on all 4 sites must have the same amplitude:
\begin{equation}
{\displaystyle \lvert\mu_A\rvert=\lvert\mu_B\rvert=\lvert\mu_C\rvert=\lvert\mu_D\rvert},\label{eq:equality}
\end{equation} 
Now we will reduce equations (\ref{eq:propA})-(\ref{eq:propD}) using (\ref{eq:equality}).
By combining (\ref{eq:propA}) and (\ref{eq:propB}):
\begin{equation}
{\displaystyle \mu_{\mathbf{k}_0}=-\mu_{\mathbf{k}_2} \textrm{ or } \mu_{\mathbf{k}_1} = -\mu_{\mathbf{k}_3}}.\label{eq:red1}
\end{equation}
By combining (\ref{eq:propA}) and (\ref{eq:propC}):
\begin{equation}
{\displaystyle \mu_{\mathbf{k}_2}=-\mu_{\mathbf{k}_3} \textrm{ or } \mu_{\mathbf{k}_1} = -\mu_{\mathbf{k}_2}}.\label{eq:red2}
\end{equation}
By combining (\ref{eq:propA}) and (\ref{eq:propD}):
\begin{equation}
{\displaystyle \mu_{\mathbf{k}_1}=-\mu_{\mathbf{k}_2} \textrm{ or } \mu_{\mathbf{k}_0} = -\mu_{\mathbf{k}_3}}.\label{eq:red3}
\end{equation}
We have 3 sets of equations (\ref{eq:red1})-(\ref{eq:red3}), each with two alternatives - that is together 8 combinations.
Only 4 of them are not in contradiction with each other and result in:
\begin{equation}
{-\mu_{\mathbf{k}_0} = \mu_{\mathbf{k}_1} = \mu_{\mathbf{k}_{2}} = \mu_{\mathbf{k}_{3}}},\label{eq:result1}
\end{equation}
\begin{equation}
{\mu_{\mathbf{k}_0} = -\mu_{\mathbf{k}_1} = \mu_{\mathbf{k}_{2}} = \mu_{\mathbf{k}_{3}}},\label{eq:result2}
\end{equation}
\begin{equation}
{\mu_{\mathbf{k}_0} = \mu_{\mathbf{k}_1} = -\mu_{\mathbf{k}_{2}} = \mu_{\mathbf{k}_{3}}},\label{eq:result3}
\end{equation}
\begin{equation}
{\mu_{\mathbf{k}_0} = \mu_{\mathbf{k}_1} = \mu_{\mathbf{k}_{2}} = -\mu_{\mathbf{k}_{3}}}.\label{eq:result4}
\end{equation}
This means that one of the moments associated with propagation vectors is always flipped. Using equations (\ref{eq:propA})-(\ref{eq:propD}), one will get:
\begin{equation}
{-\mu_{A} = \mu_{B} = \mu_{C} = \mu_{D} = 2\mu_{\mathbf{k}_0}},\label{eq:final1}
\end{equation}
\begin{equation}
{\mu_{A} = -\mu_{B} = \mu_{C} = \mu_{D} = 2\mu_{\mathbf{k}_0}},\label{eq:final2}
\end{equation}
\begin{equation}
{\mu_{A} = \mu_{B} = -\mu_{C} = \mu_{D} = 2\mu_{\mathbf{k}_0}},\label{eq:final3}
\end{equation}
\begin{equation}
{\mu_{A} = \mu_{B} = \mu_{C} = -\mu_{D} = 2\mu_{\mathbf{k}_0}}.\label{eq:final4}
\end{equation}
Equations (\ref{eq:final1})-(\ref{eq:final4}) are analogical because of the crystal symmetry.
It means, that one of the moments within the magnetic unit cell is always flipped.
For the consideration in the paper, we have chosen equation (\ref{eq:final2}).

\section{\label{sec:relfactors}Reliability factors}

We present results of the FullProf data treatment of magnetic structures in the AF1 and AF2 phase in Ho$_2$RhIn$_8$. 
Refinement was not possible for propagation vector $\mathrm{k}_1$  in vertical magnet, because all measured reflections were forbidden. 
It could not be performed for the propagation vector $\mathrm{k}_3$ either, 
because only one reflection was reachable with this propagation.
Determined magnetic moments for all phases together with the reliability factors are stated in Table~\ref{table:reliability}.

\begin{table*}[!ht]
\caption{Summary of FullProf treatment of the measured reflections in Ho$_2$RhIn$_8$.} \label{table:reliability}
 \begin{ruledtabular}
    \begin{tabular}{lccccccc}
    Phase              &  AF1 & \multicolumn{6}{c}{AF2} \\
    Type of magnet     &	- & hor. & hor. & ver. & ver. & ver. & ver. \\
    Propagation vector &  $\mathbf{k}$ & $\mathbf{k}_0$ & $\mathbf{k}_1$ & $\mathbf{k}_0$ & $\mathbf{k}_1$ & $\mathbf{k}_2$ & $\mathbf{k}_3$ \\
    \hline
    Number of reflections & 7 & 7 & 7 & 3 & 16 & 16 & 1 \\
    Magnetic moment $\mu_\mathrm{k}$ ($\mu_B$) & 6.9(2) & 3.5(2) & 3.7(2) & 3.8(6) & - & 4.0(2) & - \\
    \hline
    $RF^2$ & 17.0 & 81.7 & 9.37 & 40.4 & - & 22.8 & - \\
    $RF$   & 7.91 & 44.0 & 5.26 & 21.6 & - & 11.9 & -  \\
    $\chi$ & 0.53 & 2.09 & 0.14 & 2.86 & - & 1.06 & -  \\
    \end{tabular}
 \end{ruledtabular}
\end{table*}

\section{\label{sec:cf-fit}Crystal field data treatment}

Holmium ions occupy sites with tetragonal symmetry, 
which simplifies CEF Hamiltonian to only 5 independent parameters:
\begin{equation}
\label{HamCEF}
\mathcal{H}_{\rm CEF} = B_{2}^{0}O_{2}^{0} + B_{4}^{0}O_{4}^{0} + B_{4}^{4}O_{4}^{4} + B_{6}^{0}O_{6}^{0} + B_{6}^{4}O_{6}^{4},
\end{equation}
where $B_l^m$ are CEF parameters and $O_l^m$ are Stevens operators \cite{hutchingsCEF}.
Magnetic susceptibility $\chi$ includes CEF susceptibility $\chi_{\rm CEF}$,
the molecular field contribution $\lambda$ and already determined $\chi_0$:
\begin{equation}
\label{eq:susc}
\chi = \frac{1}{\chi^{-1}_{\rm CEF} - \lambda}+\chi_0,
\end{equation}
where $\chi_{\rm CEF}$ is given by
\begin{eqnarray}
\label{chiCEF}
\chi_{\rm CEF} = N(g_{J}\mu_{\rm B})^2 
e^{\frac{E_{n}}{k_{\rm B} T}} \nonumber\\
\left(\sum_{m \neq n} \mid \langle m \mid J \mid n \rangle
\mid^{2} \frac{1-e^{-\frac{\Delta_{m,n}}{k_{\rm B} T}}}{\Delta_{m,n}}
e^{-\frac{E_{n}}{k_{\rm B} T}} \right. \nonumber\\
\left. + \frac{1}{k_{\rm B} T} 
\sum_{n} \mid \langle n \mid J \mid n \rangle
\mid^{2} e^{-\frac{E_{n}}{k_{\rm B} T}} \right),
\end{eqnarray}
where $g_{J}$ is the Land\'{e} g-factor, $J$ is the component of the angular momentum
in direction of applied magnetic field and $\Delta_{m,n} = E_n - E_m$.

Experimental data in the whole temperature range were fitted with Eq. \ref{eq:susc}
using nonlinear optimization algorithm called
Generalized Reduced Gradient (GRG2) \cite{GRG2}. We have used
randomly selected starting parameters and run method more than 1000 times 
to find not only local, but also global minima. 

\bibliography{Ho218-field}

\end{document}